# PXIE LOW ENERGY BEAM TRANSPORT COMMISSIONING*

L. Prost#, M. Alvarez, R. Andrews, J.-P. Carneiro, B. Hanna, V. Scarpine, A. Shemyakin, Fermilab, Batavia, IL 60510, USA
R. D'Arcy, University College London, London, WC1E 6BT, UK
C. Wiesner, IAP, Goethe University, Frankfurt am Main, Germany

*Abstract*

The Proton Improvement Plan II (PIP-II) at Fermilab is a program of upgrades to the injection complex [1]. At its core is the design and construction of a CW-compatible, pulsed $H^-$ superconducting RF linac. To validate the concept of the front-end of such machine, a test accelerator (a.k.a. PXIE) is under construction [2]. It includes a 10 mA DC, 30 keV $H^-$ ion source, a 2 m-long Low Energy Beam Transport (LEBT), a 2.1 MeV CW RFQ, followed by a Medium Energy Beam Transport (MEBT) that feeds the first of 2 cryomodules increasing the beam energy to ~25 MeV, and a High Energy Beam Transport section (HEBT) that takes the beam to a dump. The ion source and LEBT, which includes 3 solenoids, several clearing electrodes/collimators and a chopping system, have been built, installed, and commissioned to full specification parameters. This report presents the outcome of our commissioning activities, including phase-space measurements at the end of the beam line under various neutralization schemes obtained by changing the electrodes' biases and chopper parameters.

## LEBT DESCRIPTION

The present layout of the PXIE beam line is shown on Figure 1. It consists of an $H^-$ Volume-Cusp Ion Source [3], capable of delivering up to 15 mA DC at 30 keV, 3 solenoids, a chopping system, Electrically Isolated Diaphragms (EID) (water-cooled, except for EID #4), an electrically isolated, water-cooled, vertical scraper assembly (a.k.a. "LEBT scraper"), and diagnostics [4]. Details about most of these components can be found in Ref. [5]. There is also a set of 4 electrically isolated scraper jaws (up, down, left and right) temporarily installed between the first two solenoids (not water-cooled). The LEBT only missing component is a switching dipole magnet which will eventually allow selecting between two ion sources (in PIP-II).

The EIDs can be biased to up to +50V in order to prevent background ions from moving from one section of the LEBT to another. Depending on the relative potential of these electrodes, different neutralization patterns may be produced. The EIDs also act as scrapers to eliminate transverse tails particles and as a diagnostic tool for beam size and position measurements [4].

The chopper consists of two parallel plates. When the beam is interrupted, -6kV is applied to the bottom ('kicker') plate, and the beam is directed to the upper ('absorber') plate. In addition to the chopping system, a modulator was added to the ion source extraction electrode. Both can produce 1 μs-16.6 ms pulses at up to 60 Hz. Thus, for commissioning of the SRF modules, the initial nominal operation scenario will be to create 5-10 μs pulses with the LEBT chopper out of milliseconds-long ion source pulses. Then, the duty factor will be increased by extending the chopped pulse as the beam line is better understood.

One key aspect of the PXIE warm front-end is its vacuum management approach. For the RFQ, it was decided to keep the vacuum level in the $10^{-7}$-$10^{-6}$ Torr range, hence limiting the gas load allowed from the LEBT. The idea is that better vacuum will improve reliability and lifetime of the RFQ, which operates CW. Also, because of the relative proximity of the first superconducting structure, it is important to limit the number of fast neutrals that could potentially reach it.

A direct consequence from this choice is the design of a long (2.2 m) LEBT with extensive pumping to effectively isolate the inherently high pressure found near the ion source from the low pressure required in the RFQ. In measurements made with a configuration similar to the one shown on Figure 1, the base pressure at the chopper assembly with the plasma on but without extracting beam is found to be < $10^{-7}$ Torr. With the entire beam being deflected onto the absorber, the vacuum remained < $2\times10^{-7}$ Torr for beam currents of up to 6 mA DC.

Maintaining a low vacuum pressure also implies long neutralization times from the background ions; therefore the beam parameters may vary dramatically throughout the pulse, which in turn can cause large beam losses. In part to alleviate this issue, an atypical transport scheme has been proposed [6], where the major portion of the LEBT is kept neutralized while positive ions are cleared in the last ~1 m of the beam line before the RFQ by applying -300V DC voltage to the kicker plate.

A drawback foreseen with this scheme was the probable emittance growth that would result from the un-neutralized transport portion of the LEBT. However, simulations that used the ion source measured parameters indicated that the final emittance would remain within the RFQ specifications. In addition, the LEBT lattice is compatible with a fully neutralized transport solution, should the scheme with the un-neutralized section be found to be too detrimental to the beam quality.

## COMMISSIONING MILESTONES

The LEBT was assembled and commissioned in phases, adding components as they were available [5].

#lprost@fnal.gov

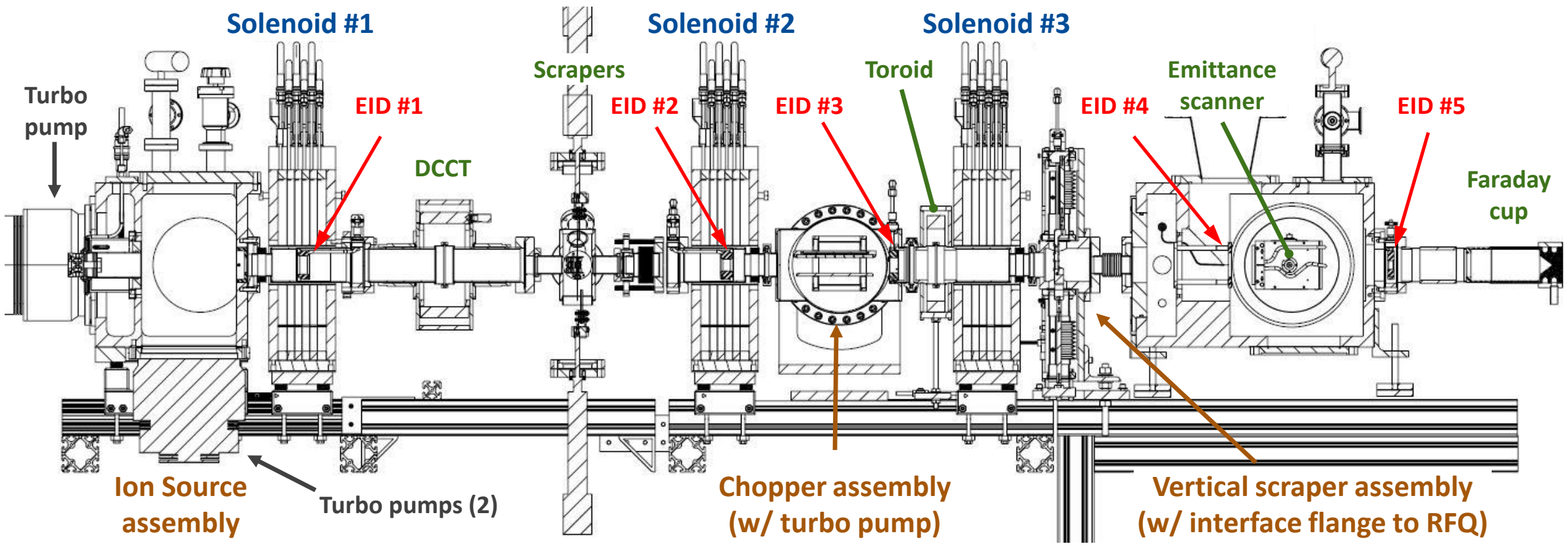

Figure 1: PXIE LEBT beam line (side view).

More than 10 mA was transported through all 3 solenoids with minimal uncontrolled losses (< 2%), both in pulsed and DC mode. At the nominal current of 5 mA, the beam Twiss parameters are close to those required for the RFQ, the emittance is low emittance (i.e. ~0.15 mm mrad, rms, normalized) and 24-hour runs without beam interruptions have been demonstrated. However, the Twiss parameters can vary significantly both during continuous running and day-to-day operation with fixed settings.

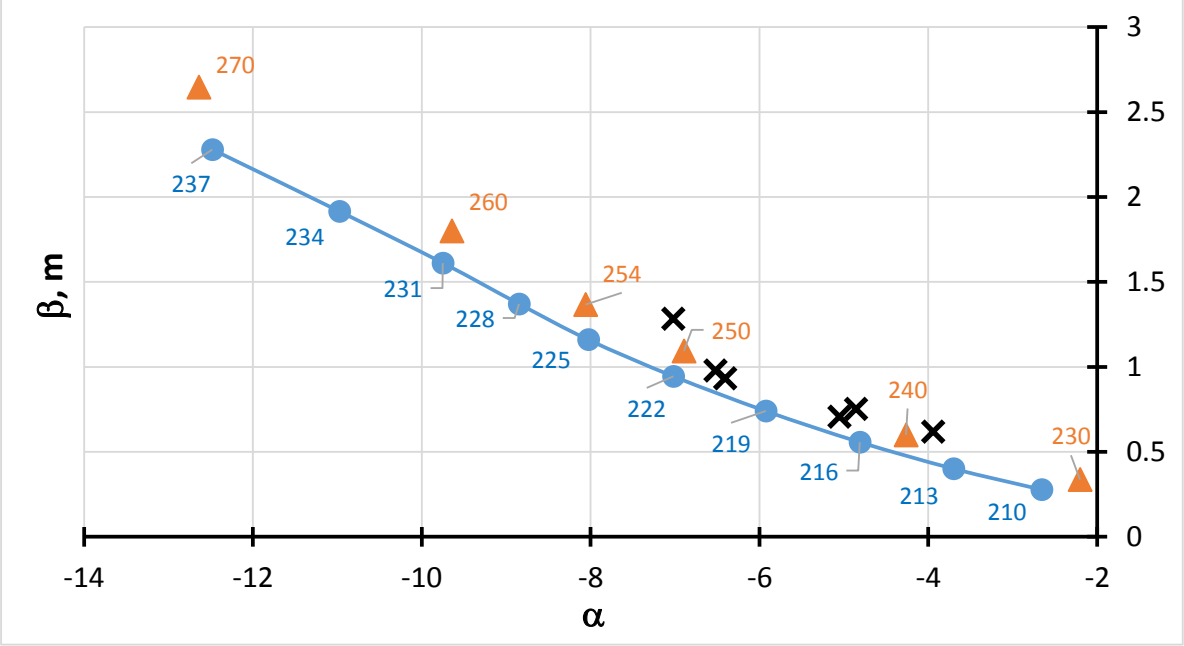

Figure 2: Twiss α vs. β at the end of the beam line. Blue circles: TRACK simulations varying Solenoid #3 current; Orange triangles: measurements for the same Solenoid #1 and #2 settings; Black crosses: day-to-day measurements with Solenoid #3 current at 240 A. 5 mA beam, 5 ms pulse chopped down to 3 ms. Data are for a time slice at the end of the chopped pulse.

While the origin of this drift has not been identified yet, it appears that merely adjusting the last solenoid current can correct for the mismatch. Figure 2 illustrates these last two points. It shows the data collected with the same machine parameters on different days compared to data from a study where the last solenoid current only was varied. The fact that the two data sets coincide should allow using the last solenoid current to maintain constant Twiss parameters over time, for instance, in a feedback loop where the current from the "LEBT scraper" would be the parameter to keep constant, since its fluctuations should indicate variations of the beam size.

## ION CLEARING

The LEBT was tested in two modes: enhanced neutralization (EID #1, #2 & #4 at +40-50V, no DC offset on the kicker electrode); and ion clearing mode (EID #1 & #2 at +40V, -300V on the kicker electrode), the latter corresponding to the baseline transport scheme.

Figure 3 shows the phase space portraits acquired with the emittance scanner for these two configurations. For the time slice displayed, which is taken near the end of the chopped pulse, the measured emittance is low and nearly the same as well as consistent with measurements of the source emittance proper (though in a different mode of operation). Thus, solutions exist for which clearing the ions does not lead to emittance growth, indicating that the proposed scheme may work. Note that although the exact level of neutralization for each case is not known, they clearly differ since the Twiss parameters are not the same, while focusing settings were unchanged.

The emittance scanner can also provide the behaviour of the beam properties with microsecond time resolution [3]. The measurements indicate that different neutralization regions and time scales exist in the LEBT, ranging from tens of μs near the ion source to ms at the end of the LEBT. For a quantitative understanding, a neutralization model is being developed. First estimations of the longitudinal potential distribution for an un-neutralized beam were performed (leading to the identification of different neutralization regions), and the longitudinal distribution of the residual gas pressure (giving the production rate of the positive ions) was simulated.

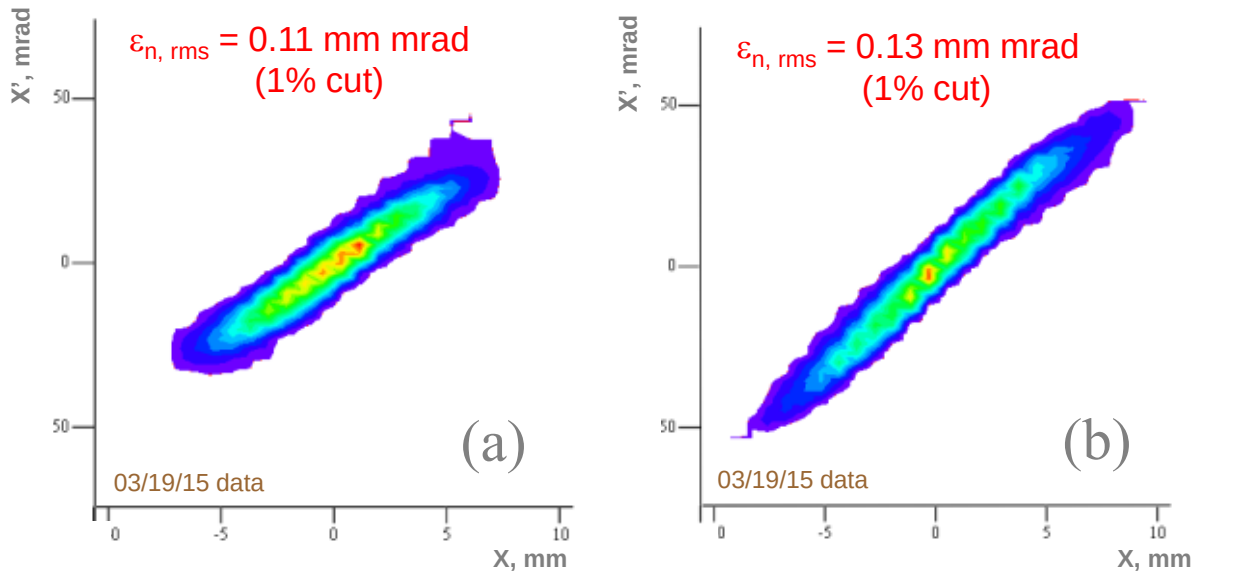

Figure 3: Phase space portraits for (a) neutralization enhanced and (b) ion clearing configurations. The background cut is 1% of the peak intensity. 5 mA beam, 5 ms pulse chopped down to 3 ms.

This first approximation of the model was validated by measuring the loss current of positive ions to the negatively biased kicker plate for different biasing schemes along the beam line. The results suggest that the degree of neutralization in front of EID #1 is limited to ~70% by the longitudinal loss of positive ions to the ion source.

## BEAM DISTRIBUTION

The low emittance growth in the transport scheme described in [6] relies on a nearly uniform beam current density distribution coming out of the ion source. Ref. [6] argues that a beam that starts with a uniform current density distribution naturally suppresses emittance growth if the transition from neutralized to un-neutralized transport takes place at the location of the image plane of the first lens, where the beam distribution is by definition the same as the object's (in the approximations of the model). Similarly, if the focal length of the lens is changed (i.e. solenoid current varied), the beam current density distribution at a fixed location should evolve and, for some solenoid current, it should be uniform.

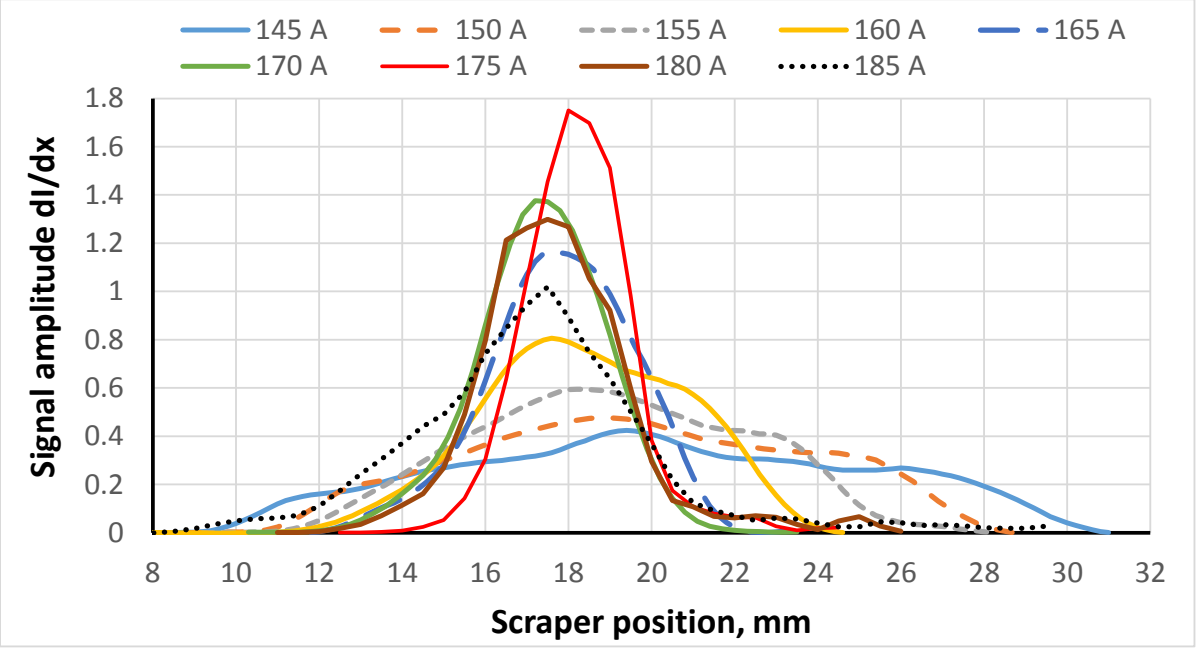

Figure 4: Ensemble of beam profiles taken with the scraper following the first solenoid (right jaw). The legend shows the Solenoid #1 currents. 5 mA beam, 1 ms pulse.

To verify the latter, profiles were acquired for several Solenoid #1 currents with the scrapers located between the first two solenoids. Figure 4 shows the reconstructed 1D profiles. Note that for a uniform current density distribution, a 1D profile is an inverted parabola, while a Gaussian distribution remains Gaussian.

While none of the profiles in Figure 4 are exactly parabolic, for the lower range of solenoid currents, they are not Gaussian either. Hence, it is likely that the beam current density distribution at the plasma surface is closer to being uniform than Gaussian, further validating the transport scheme with un-neutralized section.

## COMPARISON WITH SIMULATIONS

Figure 2 also shows the results from TRACK [7] simulations in which beam neutralization is 100% upstream of Solenoid #2 and zero downstream. The input Twiss parameters are from measurements of the ion source alone and the particles' distribution is Gaussian in both the position and velocity spaces (i.e. double-Gaussian). The current values for the last solenoid obtained with TRACK for a given ($\alpha$, $\beta$) pair do not match the solenoid currents in the experiment, although they do show a similar behaviour. The same results but expressed as the dependence of the rms beam size and emittance on the Solenoid #3 current are shown in Figure 5. After checking the solenoids' field calibrations carefully (by measuring the displacement of the beam as a function of the solenoid dipole correctors' currents), several reasons may be brought forward to explain the discrepancies. First, in Figure 5a, the fact that the curves minima do not coincide is consistent with an inaccurate representation of the neutralization pattern in the calculations. Then, the very different behaviour of the emittance between data and simulations in Figure 5b could be attributed to an incorrect description of the initial beam distribution as well as instrumental effects, which may artificially increase the measured emittance for beams with large converging/diverging angles.

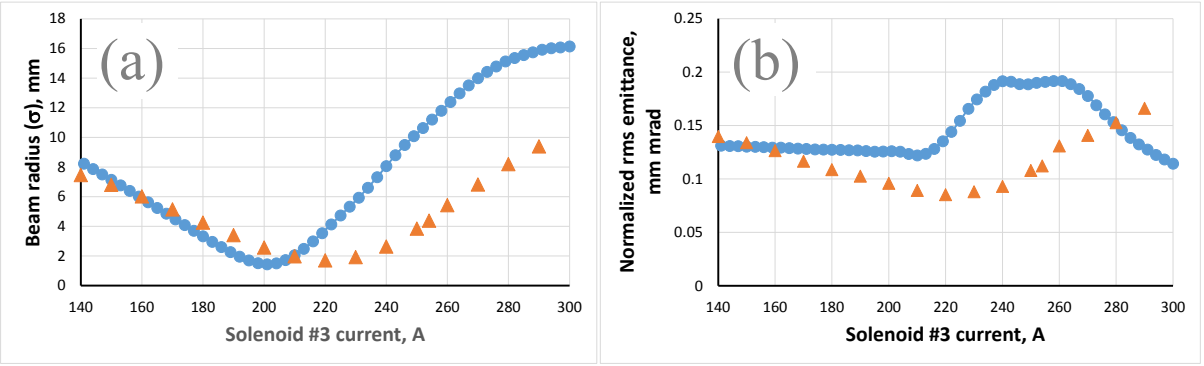

Figure 5: Beam size (a) and Emittance (b) vs Solenoid #3 current. Blue circles: TRACK simulations; Orange triangles: Measurements for the same Solenoid #1 and #2 settings. Same beam conditions as in Figure 2.

Thus, for simulations to be relied upon, more realistic models of the beam distribution and neutralization dynamics are needed. Measurements of the beam profiles after the first solenoid and other studies aimed at the understanding of neutralization along the LEBT are necessary steps toward achieving that goal.

## ACKNOWLEDGMENT

We are grateful to B. Brooker, K. Carlson, J. Czajkowski, M. Kucera, A. Saewert, G. Saewert, D. Snee, T. Zuchnik and their teams for making any of the measurements possible. Also, we acknowledge the participation of F.G. Garcia in running shifts and taking data.